\documentclass{article}
\usepackage{amssymb, amsmath, amsfonts, amsthm}
\usepackage{authblk} 
\usepackage{enumerate}
\usepackage{graphicx}
\usepackage{pdfpages}
\usepackage{multicol}
\usepackage{float}
\usepackage{caption}
\usepackage{subcaption}
\usepackage{float}
\floatstyle{plaintop}
\restylefloat{table}
\usepackage[tableposition=top]{caption}

\usepackage[backend=biber, style=authoryear]{biblatex}
\renewbibmacro{in:}{}
\newcommand{\rel}{\operatorname{rel}}
\newcommand{\tsup}{\operatorname{tsup}}
\DeclareFieldFormat{pages}{#1}
\DeclareFieldFormat[article, incollection, book, inproceedings,     unpublished]{title}{#1}
\DeclareMathOperator {\F}{FCV}
\theoremstyle{plain}

\theoremstyle{definition}
\newtheorem*{defn}{Definition}

\begin{document}

\title{Importance of overnight parameters to predict Sea Breeze on Long Island}
\author[1,3]{Kira Adaricheva}
\affil[1]{Department of Mathematics}
\author[2,3]{ Jase E. Bernhardt}
\affil[2]{Department of Geology, Environment and Sustainability}
\author[3]{Wenxin Liu}
\affil[3]{Hofstra University}
\author[3]{Briana Schmidt}

\maketitle
\begin{abstract}
    The sea breeze is a phenomenon frequently impacting Long Island, New York, especially during the spring and early summer, when land surface temperatures can exceed ocean temperatures considerably. The sea breeze influences daily weather conditions by causing a shift in wind direction and speed, limiting the maximum temperature, and occasionally serving as a trigger for precipitation and thunderstorms. Advance prediction of the presence or absence of the sea breeze for a certain location on a given day would therefore be beneficial to weather forecasters. To forecast sea breeze occurrence based on the previous night's weather conditions, we used a novel algorithm called the $D$-Basis. We analyzed sea breeze data from a recent four year period (2017-2020) at a single weather station several miles inland from the coast. High or constant station pressure, high or constant dew point, and onshore wind from the previous night were found to be strong predictors of sea breeze formation the following day. The accuracy of the prediction was around 74\% for June 2020. Unlike other prediction methods which involve the comparison of sea surface and land surface temperatures in near real time, our prediction method is based on the parameters from the prior night, allowing it to potentially aid in advanced forecasting of the sea breeze. 
\footnote{Key Words: sea breeze, weather forecasting, Long Island, $D$-basis algorithm, association rules analysis}   
\end{abstract}

\tableofcontents

\section*{Impact Statement}

The sea breeze is a phenomenon frequently influencing Long Island, especially during warm season months, due to differences in land and ocean surface temperatures. Local weather impacts from the sea breeze can include a wind shift and decreasing air temperature. Given its importance to Long Island’s climatology, a tool to forecast the presence or absence of the sea breeze in near real time is desirable. To do so, we used a novel algorithm, the $D$-Basis, to predict sea breeze occurrence during June 2017-2020, based on weather and climate data from the previous 24 hours. The algorithm was able to predict the sea breeze with relatively high accuracy and can be adapted as a tool for operational forecasters in the future. 

\section{Introduction}

Due to its location situated between the Atlantic Ocean and Long Island Sound, a sea breeze is common on Long Island during warm season months. The sea breeze originating from the south shore of Long Island, which faces the Atlantic Ocean, can be especially potent, moving several miles inland and sometimes reducing the near-surface air temperature by 5-10 degrees Celsius, see
\cite{novak2006observations}.

For example, \cite{colle2003multiseason} describes a representative event occurring on 7 June 2001, in which a sea breeze boundary moved into the middle of Long Island by the late afternoon, with a temperature gradient of 3-5 degrees Celsius across the immediate coastline. Further, that study objectively cataloged Long Island sea breeze instances during the 2000 and 2001 warm seasons, with an event required to have an approximately 6 degree Celsius temperature gradient between a near offshore buoy and a surface observing station near the coast, along with light winds during the morning. That climatology indicated that sea breeze events were by far most common in the month of June during those years, owing to a large difference between land surface and ocean temperatures typical during the late spring and early summer. More recently, \cite{McCabe2023} objectively identified Long Island sea breeze events between 2010 and 2020 using a  variety of surface, near-surface, and lower atmospheric weather observations. Those variables included near-surface temperature and moisture, sea surface temperature, station pressure, and wind speed and direction at both 10 meters and 100 meters above the surface. That study determined an average of 32 sea breeze days annually, with a maximum in July and the surrounding warm season months. 

Given the high frequency of the sea breeze on the south shore of Long Island during the early portion of the warm season, and its ability to reduce air temperature and increase wind speed in the lower atmosphere, improved advance prediction of this phenomena could be advantageous to weather forecasters. The impact of the sea breeze is further magnified by Long Island's high population density, so the ability to predict for a given day whether the sea breeze will reach a certain point inland would help with projecting energy demand, outdoor work and recreation conditions, and other societal impacts. Thus, the goal of this study is to apply the $D$-basis algorithm to the problem of short term sea breeze prediction for the south shore of Long Island. We hypothesize that by using antecedent conditions the night before a potential sea breeze event, the $D$-basis algorithm can forecast whether or not the sea breeze will reach a certain weather station on Long Island with a high degree of accuracy.

An important advance of this project compared to earlier applications of the $D$-basis, or other analysis tools involving association rules, is the development of the new methodology of processing of retrieved rules. Given the multiple runs of the algorithm and the ranking of the attributes associated with the particular outcome (in our case: either a sea breeze or non-sea breeze day), we aggregate the results and produce a unique numerical value, which determines the forecast.

\section{Data}

To objectively determine the presence of the sea breeze, 5-minute near-surface weather observations were acquired from the Hofstra University WeatherSTEM network in Nassau County, New York, specifically the  station at Hofstra's Soccer Stadium (Figures 1 and 2). Weather data at the station is recorded every minute, however, some weather observations, such as wind speed and direction, are highly variable or occasionally unavailable at that temporal resolution. Thus, all one minute observations were averaged into 5-minute intervals so that  time intervals with more readings would not weigh unequally in later groupings. 

\begin{figure}[htbp] 
\begin{center}
\includegraphics[scale=0.2]{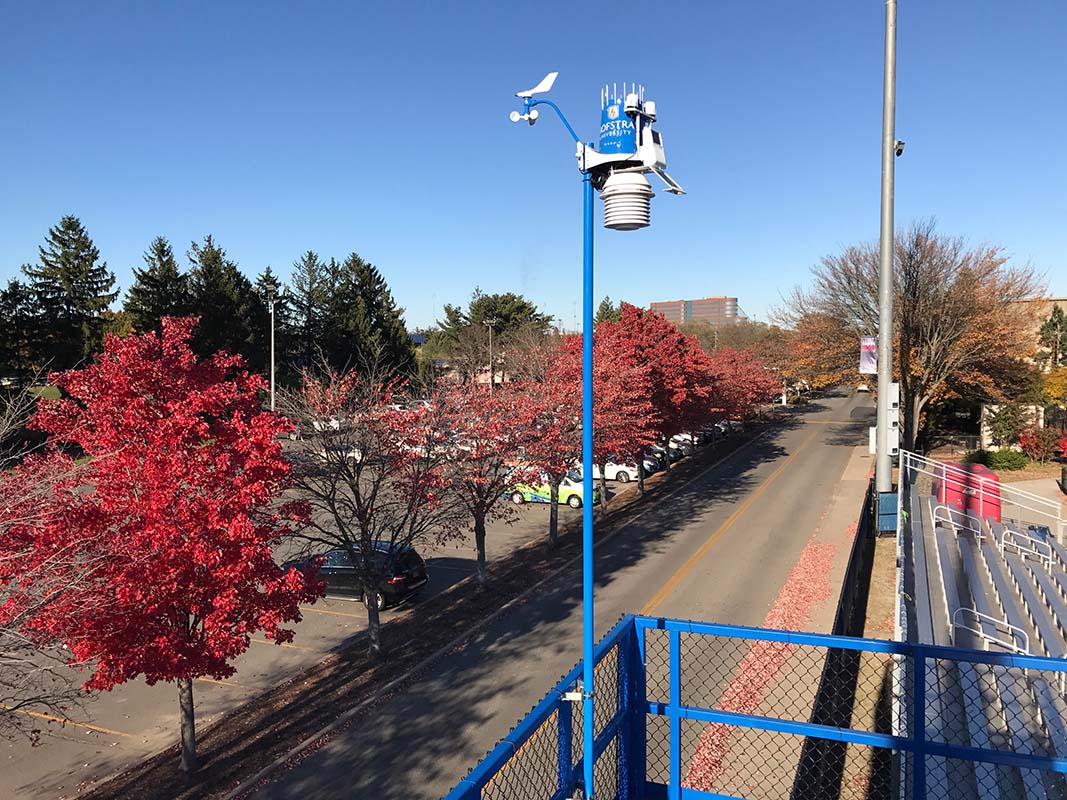}
\caption{Image of Hofstra University Soccer Staidum WeatherSTEM Station}
\label{fig:Hofstra}
\end{center}
\end{figure}

\begin{figure}[htbp] 
\begin{center}
\includegraphics[scale=0.4]{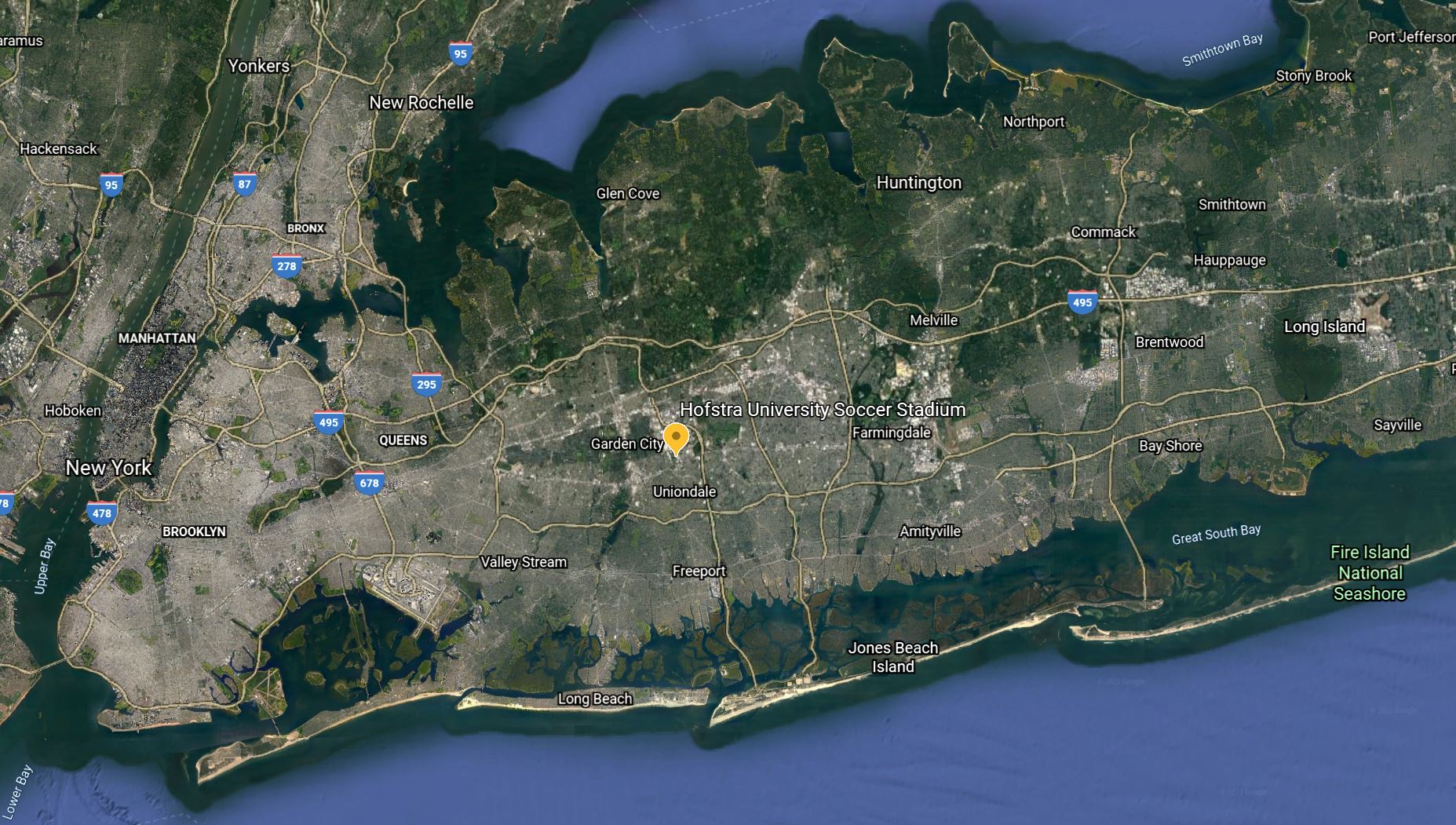}
\caption{Map showing location of the Image of Hofstra University Soccer Staidum WeatherSTEM Station, Basemap source: Google Earth}
\label{fig:hof_map}
\end{center}
\end{figure}

Weather variables most closely associated with the sea breeze were obtained for the analysis, including temperature, station pressure, dew point, and wind direction. Those data were downloaded for the months of June 2017, June 2018, June 2019, and June 2020. 

\begin{figure}[htbp] 
\begin{center}
\includegraphics[scale=0.5]{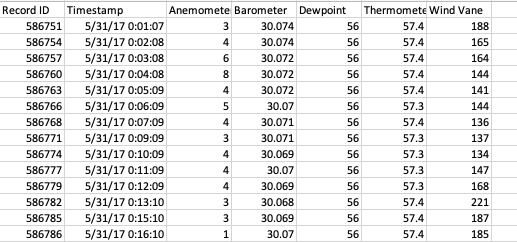}
\caption{Raw data}
\label{fig:Raw}
\end{center}
\end{figure}

Of the weather data considered, temperature and wind direction were used to objectively determine whether a particular day was a sea breeze event, while the remaining variables were included for later analysis as potential predictors in the algorithm. A candidate day was classified as a sea breeze day if the maximum temperature occurred between 10 AM and 3:30 PM local time and was followed by a period with at least 70\% of the 5-minute wind direction observations coming from an onshore direction (i.e., between 70 and 250 degrees, perpendicular to the south shore of Long Island) for the subsequent two hours.

\begin{figure}[htbp] 
\begin{center}
\includegraphics[scale=0.5]{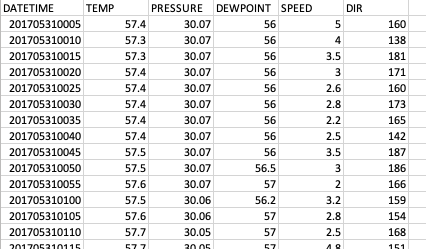}
\caption{5-minute interval data data}
\label{fig:5min}
\end{center}
\end{figure}

Those criteria were chosen based on the previous literature showing that the sea breeze boundary frequently reaches Hofstra University's location in the middle of Long Island by late afternoon. Further, once the boundary passed through Hofstra, it could be expected that the wind would come from a direction mainly perpendicular to the shore. 120 days (four months) of data were checked for the occurrence of the sea breeze, with 54 days being classified as sea breeze days, while eight were so-called 'very moist' days, as discussed later in this section,
and the remaining 58 days were categorized as non-sea breeze days.

Some days were automatically categorized as non-sea breeze days, based on the Spatial Synoptic Classification (SSC), an air mass climatology index, see \cite{sheridan2002redevelopment}. The SSC analyzes surface observations to classify the \emph{weather type} of a given day for a weather station. The nearest weather station to Hofstra University for which the SSC is available: JFK airport, in Queens, New York City --
was used, with certain \emph{weather type} days automatically being considered non-sea breeze days. Specifically, days with an SSC number of ``66" (Moist Tropical Plus) or ``67" (Moist Tropical Double Plus) were classified as \emph{very moist days} and were not eligible to be categorized as sea breeze. This was due to the fact that when very tropical air masses are present over Long Island, they are almost always accompanied by deep synoptic flow from the south or southwest, the same direction as the sea breeze. Thus, it is rare for a meaningful sea breeze boundary to develop on those days, since the larger scale flow is already coming from a similar direction. 

After the classification of \emph{very moist days} was completed, the remaining sea breeze and non-sea breeze days were investigated to determine whether the $D$-basis algorithm and chosen predictors could successfully forecast the presence or absence of the sea breeze using data from the preceding night. 

\section{Detailing the definition of sea breeze}\label{SBDef}

We began with the aforementioned definition of a sea breeze day being a day where \emph{the highest temperature occurs between 10 am and 3:30 pm} in locations near the coast. The goal of the algorithm development was to specify which other weather variables could provide advanced notice of an impending sea breeze day. Doing so could allow those weather parameters to be leveraged using the computer algorithm to forecast the sea breeze in advance. 
Indeed, while there is no universally accepted numerical definition of the sea breeze for a given locality, having a binary definition (i.e., yes sea breeze or no sea breeze) was necessary to test the algorithm developed.
Moreover, it is easier to remove false positives than to account for false negatives in developing the algorithm, so the definition of the sea breeze was slightly expanded after initial testing, to allow for more days to be classified as potential sea breeze days.
For example, the high temperature cutoff was initially set at 3 PM, but later moved back slightly to 3:30 PM.

One method used to filter out false positives was the implementation of a wind direction requirement into the sea breeze definition. Since the Atlantic Ocean is roughly due south of Long Island, the five-minute averaged wind direction in the two hours following the maximum temperature being recorded was required to be from the southern half of the wind rose (i.e., between 70 and 250 degrees) for at least 70\% of the time. That requirement helped to eliminate false positives stemming from instances such as a strong cold frontal passage, which could result in the high temperature occurring much earlier than normal during the day, but due to a reason other than the sea breeze. On the other hand, the definition should also not be too specific to a location or time of year so that it does not work as accurately in other settings. One way to mitigate against that issue was to use changes in weather variables, instead of absolute values, though the use of the latter was still necessary in relation to air pressure. Thus, the following was used as the basic definition for a sea breeze:

\begin{defn}{Sea Breeze}
\begin{itemize}
    \item[(1)] Occurs  in a coastal area.

\item[(2)] Occurs in the warm season, when the land temperature is sufficiently warmer than the water temperature.

\item[(3)] Occurs in absence of larger scale weather phenomena.

\item[(4)] Occurs near surface temperature of the day happens between 10 AM and 3:30 PM (local time, not precisely generalizable to all locations).
 
\item[(5)] For two hours after the highest temperature, the wind direction must be coming from the direction of the body of water for at least 70\% of the time (i.e., roughly perpendicular to shore).
\end{itemize}
\end{defn}

\section{Approach}

\subsection{Parameter Selection}\label{attr}

According to a formula presented in \cite{BiGr62}, the lake breeze forecasting index $\epsilon$ is proportional to the square near surface wind speed $|U|$ and inversely proportional to the land-sea temperature contrast $\Delta T$:

\[
\epsilon = \frac{|U|^2}{C_p\Delta T},
\]
where $C_p$ is a heat coefficient of dry air at constant pressure. $\Delta T = T_{land}-T_{sea}$ measures the temperature difference, with $T_{land}$ taken at the point far enough inland. Many other forecasting mechanisms are reported in \cite{MiEtAl03}, where the wind and temperature measurements were of central importance. Most of their forecasting relied on the morning measurement to predict the sea breeze happening on the same day. 

In our study, we relied solely on measurements produced by a Hofstra University weather station, located considerably inland compared to south shore of Long Island, and not sea surface temperature observations. We compared the data of one of the years with similar measurements of another weather station, near Wantagh, New York, much closer to the shore, but the difference was insignificant.

The prediction method we developed involved the examination of station pressure, dew point, and wind direction during the prior night from 7pm to 7am on the morning of the predicted day. Initial focus was on the dynamic of pressure and dew point: rising, dropping or constant, and at later stage of the project the absolute values of pressure were added to the list of attributes. 

According to \cite{LaKr01} three quarters of sea breeze events happened with cross shore component $|U_x|$ being less than 2 m/sec, thus, wind speed might not have considerable effect on the sea breeze. This attribute was not considered in current study, especially given the relative distance of the weather station from the shore. Pressure, pressure dynamic, dew point dynamic and wind direction were chosen as predictor variables based on the weather variables most closely associated with sea breeze days for Long Island.  It was also important 
to consider the development of those variables overnight, prior to sea breeze initiation, and we used 2-hour intervals when observing the dynamics of changes in parameters. Those overnight parameters, one for each 2-hour interval between 7pm and 7am prior to day of prediction, were classified as follows: 

\begin{itemize}
    \item pressure rising or dropping during the interval;
    \item dew point rising or dropping during the interval; 
    \item onshore wind direction (70-250 degree) dominating ($> 60\%$) during the interval;
    \item station pressure above or below the climatological average sea-level pressure of 29.92 inches mercury (1013 mb) during the interval (given Long Island's elevation very close to sea level).
\end{itemize}

Generally, the sea breeze occurs in June in the absence of a strong synoptic low pressure system (pressure not dropping) or cold frontal passage (dew point not dropping). On the mesoscale, persistence was also selected as an indicator for sea breeze through the onshore wind requirement. At the synoptic scale, the SSC weather type was used to remove days where a sea breeze would not be expected to occur (i.e., due to persistent onshore caused by a synoptic weather system). 

\subsection{$D$-basis algorithm and ranking of attributes by the relevance}

In this work we explored an approach in weather forecasting leveraging the discovery of association rules in the binary data. One of recent explorations of this sort was done in \cite{Coul21}.

The $D$-basis is a new algorithm described in Adaricheva and Nation 2017 that  discovers  the  \emph{implications} $S\to d$ in  a  table  with  entries  0  and  1. 
This algorithm belongs to the family of analytic tools based on \emph{association rules} and/or \emph{implications}, but it also involves secondary statistical analysis of the retrieved rules.

In the current study, $S$ is a subset of attributes/columns (weather observation attributes) and $d$ is another column (e.g.,  an  indicator  of sea  breeze  day). Note that non-binary attributes can be expressed as a combination of binary ones. For example, attributes with five values can be given by five binary columns. Attributes that have values of real numbers within some interval can be discretized by splitting the range into a few sub-intervals and assigning a new attribute to each subinterval of values. The  rows  of  the  table represent the days of the observation. Implications are association rules that hold in all rows of the table. In our context, for every day/row, if all attributes in set $S$ occur (marked by 1 in that row), the attribute $d$ occurs as well.

The algorithm is controlled by several parameters, which filter the retrieved rules $S\to d$ with the fixed target attribute $d$. The most important are the \emph{minimal support} and \emph{row deletion}. 
We say that an observation (a row of the table) \emph{validates} rule $S\to d$, when all attributes in $S$ as well as $d$ are present in the observation, that is, the entries in the row are marked by 1, corresponding to columns in $S$ and $d$. The \emph{support of the rule} is the number of observations/rows where the rule is validated. When the test runs with parameter of \emph{minimum support} $=k$, only the rules with the support of at least $k$ are returned. Further discussion of the parameters of the test is given in appendix section \ref{A:rank}.

Using the sets of rules retrieved on the two runs of the algorithm: one on the target attribute $d$, and another on its negation $\neg d$, which is not necessarily present in the data, so it needs to be created - permits computation of a real non-negative number assigned to each attribute $a$ different from $d$. It is called the \emph{relevance} of $a$ with respect to $d$ and denoted $rel_d(a)$. The higher $rel_d(a)$, the more frequently attribute $a$ appears in set $S$ for rules $S\to d$ compared to rules $S\to \neg d$. All attributes different from $d$, therefore, may be ranked by the relevance with respect to $d$, and our method would investigate the attributes with highest ranks with respect to $d=$\emph{sea breeze day} or $d=$\emph{non-sea breeze day}, which we call \emph{normal day} for the rest of the paper. 

Note that the relevance is computed within the sets of retrieved rules, one for target $d$ and another for target $\neg d$. Changing the parameters of the tests will change the sets of rules, which may change the relevance. For this reason we run multiple tests, then aggregate the results. More detailed description of the $D$-basis algorithm is given in appendix section \ref{A:Dbas}, and the definition and computation of the relevance in section \ref{A:relevance}.

\subsection{Code development for the data conversion}
The $D$-basis code was developed in github https://gitlab.com/npar/dbasis
and was first made publicly available at the time of publication in \cite{NCSLA21}. The weather data conversion for the $D$-basis entry was performed in the R programming language following the rules (3)-(5) of the Sea Breeze Definition in section \ref{SBDef}. 
The files related to this project are located in https://gitlab.com/npar/seabreeze

\section{Data conversion}\label{Data}

Since the $D$-basis algorithm acts on binary data, the weather variables collected from the Hofstra weather station were converted into binary form. Each day in the converted data was represented by a row of the table, while the columns represented a weather attribute which was marked by 1 when it was observed on a particular day/row, and by 0 when the attribute was not observed. Several attributes were the target attributes representing whether or not a sea breeze was observed, or if the day was excluded due to the synoptic climatology (i.e., a very moist day):\\ \\
71: Next Day Sea Breeze (1=SB, 0=Normal or Very Moist)\\
72: Complement of 71 (1=Normal or Very Moist, 0=SB)\\ \\
73: Next Day Normal Day (1=Normal, 0=SB or Very Moist)\\
74: Complement of 73 (1=SB or Very Moist, 0=Normal)\\

For example, the first row of the table has entry $1$ in column 71, because on June 1, 2017 a sea breeze was observed (and it also had 0 in column 72 and 73 and 1 in column 74).
Note that ``Next Day" applies to June 1, because the observation of some weather attributes were made on May 31, 2017.

The rows in the converted data were sorted as follows:\\
Rows	1-30: May 31 2017-June 29 2017\\
Rows	31-60: May 31 2018-June 29 2018\\
Rows	61-90: May 31 2019-June 29 2019\\
Rows	91-120: May 31 2020-June 29 2020\\

The first 6 attributes given in Table \ref{Tab1} were possible descriptors of the day prior to prediction, based on the SSC categorization of that day.

\begin{table}[t]
\begin{center}
\begin{tabular}{|c|c|c|}
    \hline
 Column \# & \ Column description & Entry values \\
    \hline
    1 & Sea Breeze &  1=yes, 0=no\\
    \hline
    2 & Normal Day & 1=yes, 0=no \\
    \hline
    3 & Very Moist Days & 1=yes, 0=no \\
    \hline
    4 & SSC Dry Day & 1=yes, 0=no \\
    \hline
    5 & SSC Moist Day & 1=yes, 0=no \\
    \hline
    6 & SSC Transition Day & 1=yes, 0=no \\
    \hline
\end{tabular}
\end{center}
\caption{Attributes for Day Categories}
\label{Tab1}
\end{table}

For example, the first row had entry 1 in column 2 and 0 entry in columns 1 and 3-6, because May 31, 2017, was classified as a Normal Day (i.e., no sea breeze). 

The attributes 7-16, as in Table \ref{Tab2}, measured weather parameters for the time interval 7-9 pm the preivous evening, with entry value 1=yes or 0=no.

\begin{table}
\begin{center}
\begin{tabular}{|c|c|c|}
    \hline
 Col \# & \ Column description 
 & Details \\
    \hline
    7 & Pressure rising 
    & Pressure increased by the end of interval\\
    \hline
    8 & Pressure falling 
    & Pressure decreased by the end of interval\\
    \hline
    9 & Pressure consistent 
    & No change in pressure\\
    \hline
    10 & Wind North 
    & at least 60\% of the time interval\\
    \hline
    11 & Wind South 
    &  at least 60\% of the time interval \\
    \hline
    12 & Dew point rising 
    & Dew point increased by more than 1 degree F\\
    \hline
    13 & Dew point falling 
    & Dew point decreased by more than 1 degree F \\
    \hline
    14 & Dew point consistent 
    & Dew point changed by less than 1 degree F\\
    \hline
    15 & High pressure 
    & Pressure above 29.92 in Hg
    \\
    \hline
    16 & Low pressure 
    & Pressure below 29.92 in Hg
    \\
    \hline
\end{tabular}
\end{center}
\caption{Attributes of weather observations}
\label{Tab2}
\end{table}

The pressure trend and dew point trend of each 2-hour interval were determined simply by subtracting the end value from the start value. For pressure a difference of 0.00 would designate it as ``consistent", while for dew point, a change was classified to have occurred when the difference was in interval (-1,1).

Pressure above 29.92 in Hg was categorized when all 5-minute intervals during the 2-hour period were above or on that climatological threshold; otherwise, when any 5-minute interval during the 2-hour period is dropped below the threshold, it is determined as ``Pressure below 29.92 in Hg."

Similar attributes, in the same sequence of 10, described the measurements during the following two-hour time intervals:\\
Columns 17-26  between 9 - 11 pm,\\
Columns 27-36 between 11 pm - 1 am,\\
Columns 37-46 between 1 - 3 am,\\
Columns 47-56 between 3-5 am, and\\
Columns 57-66 between 5-7 am.\\

Columns 67-70, as described in Table \ref{Tab3}, were added to further confirm the initial hypothesis that high pressure, together with non-decreasing pressure, may be a significant factor in sea breeze prediction, given that a sea breeze is likely to form in the absence of strong fronts or cyclones, which may be indicated by low or falling pressure. Hence, some combinations of the six 2-hour intervals were introduced. If 4 or more intervals possessed non-falling pressure, column 67 of that day was marked as 1. If 4 or more intervals had no time instance with pressure lower than 29.92 inches Hg, column $68$ was marked as 1.

\begin{table}[t]
\begin{center}
\begin{tabular}{|c|c|}
    \hline
 Col \# & \ Column description  
 \\
    \hline
    67 & When $>3$ of 6 intervals have rising or constant pressure 
    \\
    \hline
    68 & When $>3$ of 6 intervals have pressure above 29.92 inches Hg 
    \\
    \hline
    69 & Pressure rising or high pressure 
    \\
    \hline
    70 & Pressure falling or low pressure 
    \\
    \hline
\end{tabular}
\caption{High pressure attributes}
\label{Tab3}
\end{center}
\end{table}

These observations associated with anticyclonic conditions were selected for inclusion because they imply weak flow at the surface and aloft. Such conditions have been shown to be a key indicator of sea breezes along the US East Coast in previous studies, such as \cite{Hughes2018} and \cite{Cetola1997}.

\section{Results}

\subsection{Initial stage of the project}

The initial phase of this project was to conduct $D$-basis testing on 3 years of June data ranging from 2017-2019 to verify the effectiveness of the initial determinants in predicting the presence or absence of the sea breeze. Three  attributes were tested: the change in pressure, the change in dew point, and the dominant wind direction.

After computing the relevance, the variables that were most important for predicting a \emph{sea breeze} day were \emph{pressure rising, winds from the south (i.e., on-shore), and dew point either rising or staying consistent}. The attributes that were important for predicting \emph{normal} days were \emph{pressure falling, wind from the north (offshore), and dew point falling}.

The following tables collect the results on each weather variable with respect to 6 time intervals of 2 hours between 7pm and 7am.  Table \ref{Tab4} shows, for example, that all 6 time intervals are important for the pressure rising attribute because they are included in the top 17 (out of 58 total) attributes ranked by the relevance, when listed in order from highest to lower. 

\begin{table}[t]
\begin{center}
\begin{tabular}{|c|c|c|}
    \hline
  \ Sea breeze forecast   & Relevance & \ Rank \\
    \hline
  \  Attributes:     & \# of intervals & in the top \\
  \hline
    Pressure rising & 6 & 17 \\
    \hline
    Wind South & 6 & 24 \\
    \hline
    Dew point rising & 3 & 22 \\
    \hline
    Dew point consistent & 5 & 28 \\
    \hline
\end{tabular}
\end{center}
\caption{Number of time intervals in the sea breeze forecast}
\label{Tab4}
\end{table}

The time intervals that are not in the top ranked for southerly wind are always earlier in the night.  
This implies that southerly winds in the later hours of the night prior to sea breeze development are more consistently important. 
The combination of dew point rising and dew point consistent for prediction of the sea breeze covered all time intervals.

\begin{table}[t]
\begin{center}
\begin{tabular}{|c|c|c|}
    \hline
  \ Normal day forecast   & Relevance & \  Rank \\
    \hline
  \ Attributes:      & \# of intervals & in the top\\
  \hline
    Pressure falling & 5 & 20 \\
    \hline
    Wind North & 4 & 13 \\
    \hline
    Dew point falling & 5 & 24\\
    \hline
\end{tabular}
\end{center}
\caption{Number of time intervals in the normal day forecast}
\label{Tab5}
\end{table}

Results of predicting a normal day are in Table \ref{Tab5}.
The earlier hours of the prediction interval were more important for pressure and dew point, while the later hours were more important for the wind when predicting a normal day.

\subsection{Second stage of the project}\label{second stage}

The strong separation of patterns of the winds from the south (north), pressure rising (falling) and dew point not falling (falling) when occurring prior to a sea breeze (normal) day served as the basis to formulate the prediction method in the second stage of the project.

The main new features in the second phase of the project were:
\begin{itemize}
    \item[(1)] adding absolute values of pressure above/below the climatological threshold of 29.92" Hg (one standard atmosphere), during six 2-hours intervals;
    \item [(2)] correcting wind direction, given that the shoreline is not in a perfect west to east orientation;
   
    \item [(3)] adding 30 more observation days for June 2020;
    \item[(4)] running $D$-basis with minimum support = 6,8,11. 
\end{itemize}

\begin{figure}\label{fig:Press}
\begin{center}
    \begin{subfigure}[a]{0.4\textwidth}
        \centering
      \includegraphics[width=\textwidth]{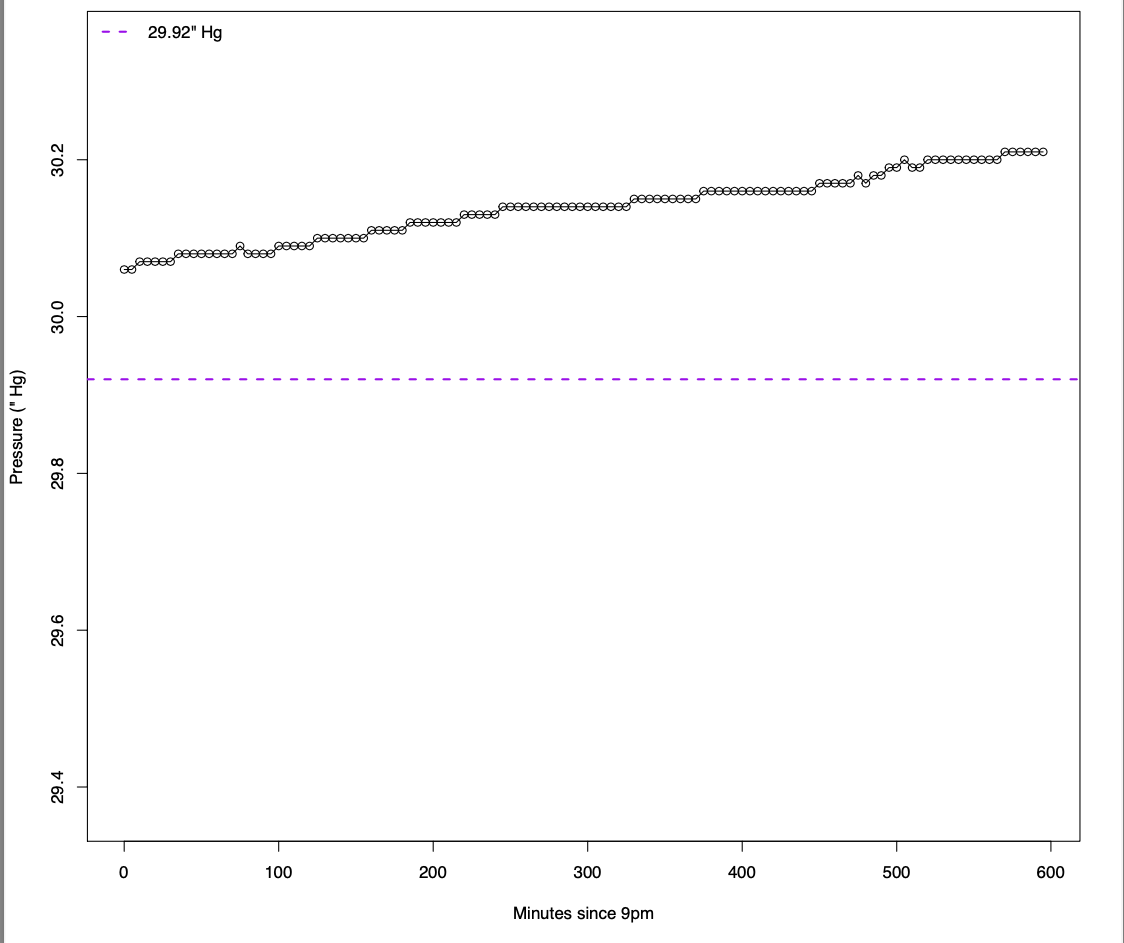}
      \caption{A typical sea breeze day pressure trend}
           \label{fig:ch1}
    \end{subfigure}
\hspace{1cm}
    \begin{subfigure}[a]{0.4\textwidth}
        \centering
        \includegraphics[width=\textwidth]{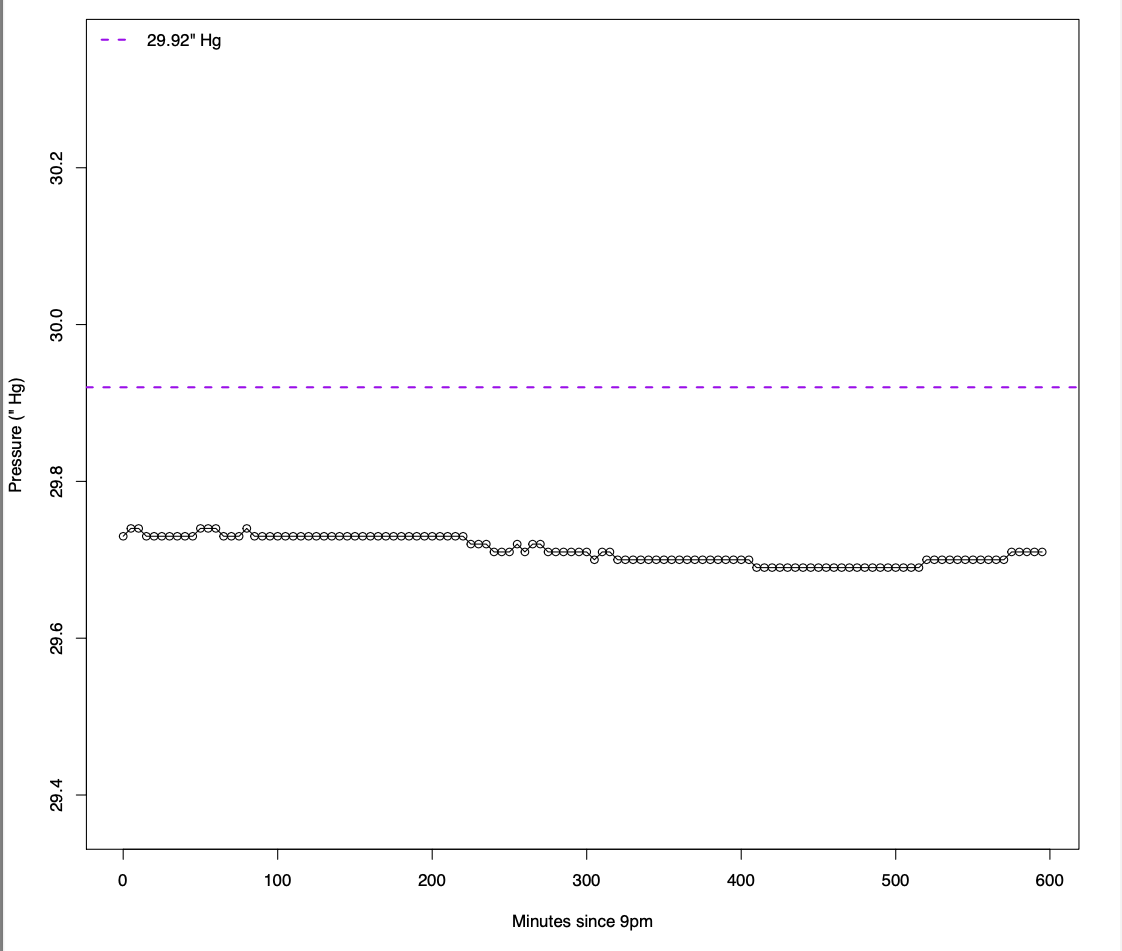}
        \caption{A typical normal day pressure trend}
            \label{fig:ch2}
    \end{subfigure}
\end{center}
\caption{Patterns of pressure measurement between 9 pm and 7am}
\end{figure}

The decision to include the absolute pressure as an additional parameter came after careful observation of the graphs of pressure behavior over 120 days. 
To examine the pressure trend for sea breeze days, one can observe the daily pressure charts.
Each dot on the daily trend graph represents a 5 minute interval during the 9pm to 7am time frame, where the purple dashed line represent the 29.92'' Hg standard that we later decided to use to determine whether each 2 hour interval is considered as \emph{high pressure}, Figures \ref{fig:ch1} and \ref{fig:ch2}.

Two pressure trend diagrams are chosen to demonstrate the general trend for a sea breeze day versus a normal day. 
On Figure \ref{fig:ch1} there is an example of  
overnight pressure chart between June 25--26 in 2018, with June 26 being a sea breeze day.

On Figure \ref{fig:ch2} there is an example of overnight pressure chart between June 20--21, 2018, with June 21 being a normal day.

The normal day trend is sometimes seen slightly decreasing/flat or increasing, but rarely remains completely above the purple dashed line. 

\begin{figure}[H] 
\begin{center}
\includegraphics[scale=0.4]{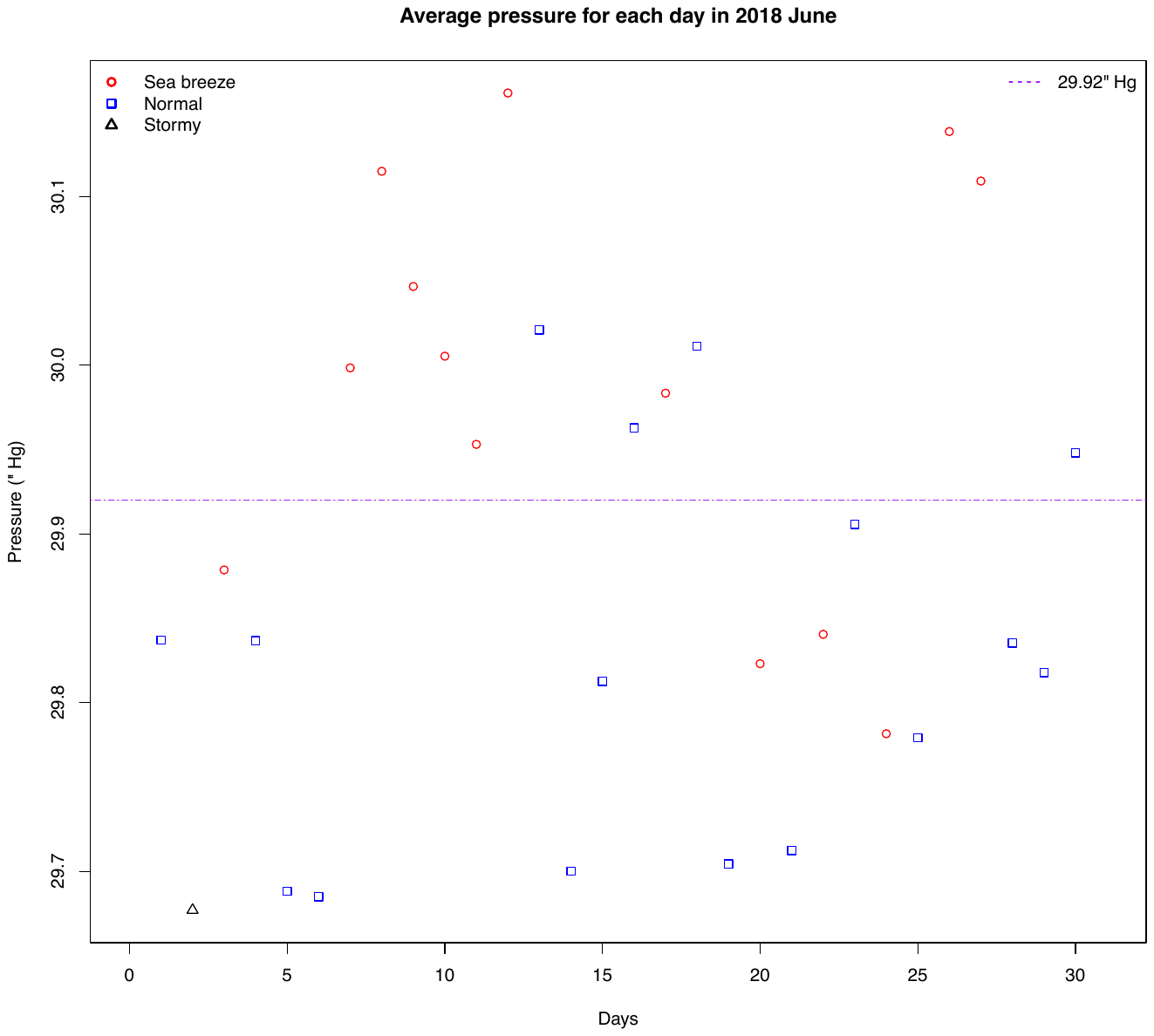}
\caption{June 2018 average pressure dot graph}
\label{fig:J2018}
\end{center}
\end{figure}

 June 2018 is a good example to account for the 29.92" Hg pressure standard. In the graph on Figure \ref{fig:J2018}, each dot represents the average pressure of each day in June from 9pm--7am; the sea breeze days (red dots/circles) are mostly above the purple dash line, while the normal days (blue dots/squares) and very moist days (black dots/triangles) are mostly below the purple line with few exceptions. 

As was observed, the two distinctive trends for normal days and sea breeze days deserve potentially more quantitative validation. Hence, we also created two more columns combining the absolute pressure values and the trend. To consolidate the 6 intervals, we created column 68 for when more than half of the 6 intervals have high pressure. Also to support both increasing and constant pressure, we created column 67 that returns 1 if at least 4 intervals have pressure rising or constant pressure. Lastly, column 69 is created as [67 OR 68] to consolidate the attributes of high pressure, pressure rising as a possible predictor of a sea breeze day.

Similarly, column 70 for possible prediction of normal days was defined as [$\neg$ 67 OR $\neg$68], so it would mark a day 1 when at least 4 intervals have low pressure or falling pressure on that day, see Table \ref{Tab3}. 

\begin{figure}[htbp] 
\begin{center}
\includegraphics[scale=0.5]{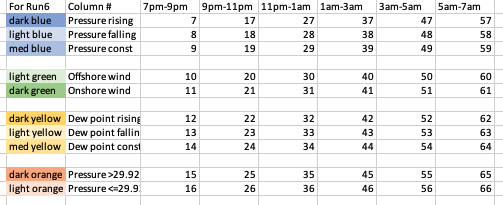}
\caption{Color coding of attributes}
\label{fig:Color}
\end{center}
\end{figure}

In addition, a color code scheme for several groups of parameters was designed to better visually represent the relevance of the attributes. For each $D$-basis run, the top 15 attributes of the highest relevance with their relevance value listed on the side were included. Those attribute numbers were then classified into ten colors based on the same attribute for different time periods. The color blue showed attributes for pressure, green for wind direction, yellow for dew point, and orange for the absolute pressure. 
Figure \ref{fig:Color} below indicates the mapping of the ten colors to column numbers with the time periods specified. 
Coincidentally, there are ten parameters for each 2-hour interval, which enables the column numbers for each parameters to be exactly ten numbers apart.

\begin{figure}[H] 
\begin{center}
\includegraphics[scale=0.5]{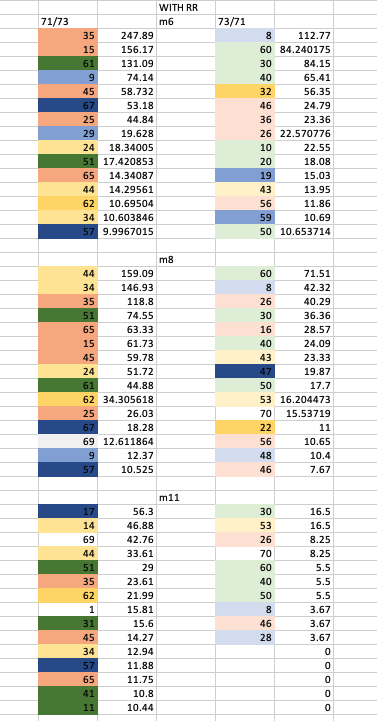}
\caption{Color coded aggregated relevance from the test with row removal}
\label{fig:Run6RR}
\end{center}
\end{figure}

\begin{figure}[H] 
\begin{center}
\includegraphics[scale=0.5]{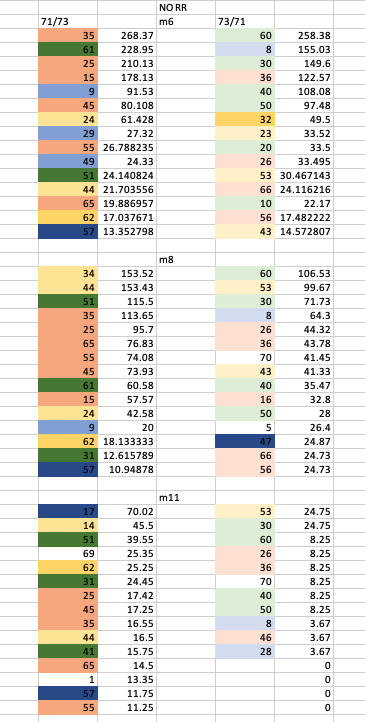}
\caption{Color coded aggregated relevance from the test without row removal}
\label{fig:Run6NoRR}
\end{center}
\end{figure}

The color scheme described above was then applied to the top 15 attributes with highest relevance, computed at minimal support=6,8 and 11, with row removal. The left side of Figure \ref{fig:Run6RR} lists the top attributes, together with normalized values of relevance, when target is the Sea breeze, and the right side is the relevance when the the target is the Normal day. Similar results are shown for the tests without row removal on Figure \ref{fig:Run6NoRR}.

The prevalence of dark colors on the left and light colors on the right demonstrates a clearly observable pattern. For the normal day prediction on the right side, most columns are of lighter colors except for one dark blue and several medium yellow. Generally, the color coded view confirms the potential predictors for both sea breeze days and normal days.

It is also observed that the columns 69 and 70 both appear in Figures \ref{fig:Run6RR} and  \ref{fig:Run6NoRR}. 

Column 69 even ranked in the top 3 relevant to sea breeze, when testing with minimal support of 11
in both graphs. On the other hand, column 70 appeared with the high rank for normal day, when testing with minimal support 8 and 11 in both graphs.

Since the color-coded relevance charts clearly demonstrate the importance of the dark colored attributes in predicting sea breeze, we attempted to incorporate those numbers into our prediction model to achieve better prediction results. 

\subsection{Computation of the Forecasting Value}

In this section we describe a novel approach to use leading attributes and their $D$-basis based relevances to compute a numerical Forecasting Value (FCV) which would predict either a sea breeze or normal day using measurements from the Hofstra weather station taken between 7pm on the previous day until 7am of the forecast day, which is a generalizable approach that could be developed for any coastal area by adjusting the coefficients of the formula.

If we choose attributes $a_1,a_2,\dots, a_s$ as the most relevant for the prediction of sea breeze, and use relevances for these attributes when targeting $d=$`sea breeze day', we determine coefficients $k_1,k_2,\dots k_s$. Similarly, we can choose $a_{s+1},a_{s+2},\dots a_{2s}$ as the most relevant attributes in predicting a normal day, and their relevances when targeting $d=$`normal day' are converted into coefficients $k_{s+1},k_{s+2},k_{2s}$. This yields a formula for computation of $\F$:
\[
\F(\overline{x})= k_1*x_1+k_2*x_2+\dots+k_s*x_s - k_{s+1}*x_{s+1} - k_{s+2}*x_{s+2}-\dots - k_{2s}*x_{2s}
\]
Here $\overline{x}=\langle x_1,x_2,\dots,x_{2s}\rangle$ is a vector describing attributes for day $x$: $x_i=1$, if attribute $a_i$ is observed on the day $x$, and $x_i=0$ otherwise. We forecast that $x$ is a sea breeze day, when $\F(\overline{x})>0$, and a normal day otherwise. 

What follows is the proposed technique of the choice of attributes and computation of the corresponding coefficients $k_1,\dots ,k_{2s}$. We chose to compute these coefficients based on relevance values given by the $D$-basis. 

We varied the minimum support in several runs of the $D$-basis: when minimal support is lower, more implications are returned; when minimal support is increased, there are less implications, but they are more valuable as they are manifested in more observations. 

The minimal support was chosen at levels 6, 8 and 11, thus, the highest support was corresponding to 10\% of the days of the observation. We also found that running $D$-basis with the row removal, that is, partly removing the rows corresponding to the very moist days, provides better prediction values. 

After all attributes were ranked by the relevance in each run, the relevance values were normalized by setting the largest relevance value as the standard 1 so that all normalized relevance values were placed in intervals of values between 0 and 1. This way, a standard measure was available for comparison across $D$-basis runs of different target columns and minimal supports. Then, for each attribute, the normalized relevance values were averaged over the three runs with different minimum supports, where more weight  was given to the values from the runs of larger minimum support. To demonstrate the effectiveness of the normalized relevance values, we composed a dot graph with their averaged values as the yellow line graph. All dots in Figure 10 are within the 0 to 1 range on the vertical axis. The yellow dots at the tops of peaks of line graph 
represent 6:8:11 ratio averages
of the relevance numbers from three runs of algorithm with different minimal supports.

More precisely, if $Rel(6,a)$, $Rel(8,a)$ and $Rel(11,a)$ are normalized relevance values for attribute $a$, when targeting, for example, the sea breeze day in the $D$-basis run with minimum support 6, 8 and 11, respectively, then the average value of relevance for this attribute would be computed with the following formula:
\[
rel(a)=\frac{6*Rel(6,a)+8*Rel(8,a)+11*Rel(11,a)}{25}
\]

These values would produce a new ranking of the attributes, one for sea breeze target, and another for the normal day target. Then subset $a_1,\dots, a_s$ of the top $s$ attributes was chosen from the sea breeze rank, and another subset $a_{s+1},\dots, a_{2s}$ was chosen from the rank of the normal day. The initial choice was $s=10$, but we also considered smaller values of $s$.\\

\begin{figure}[H] 
\begin{center}
\includegraphics[scale=0.4]{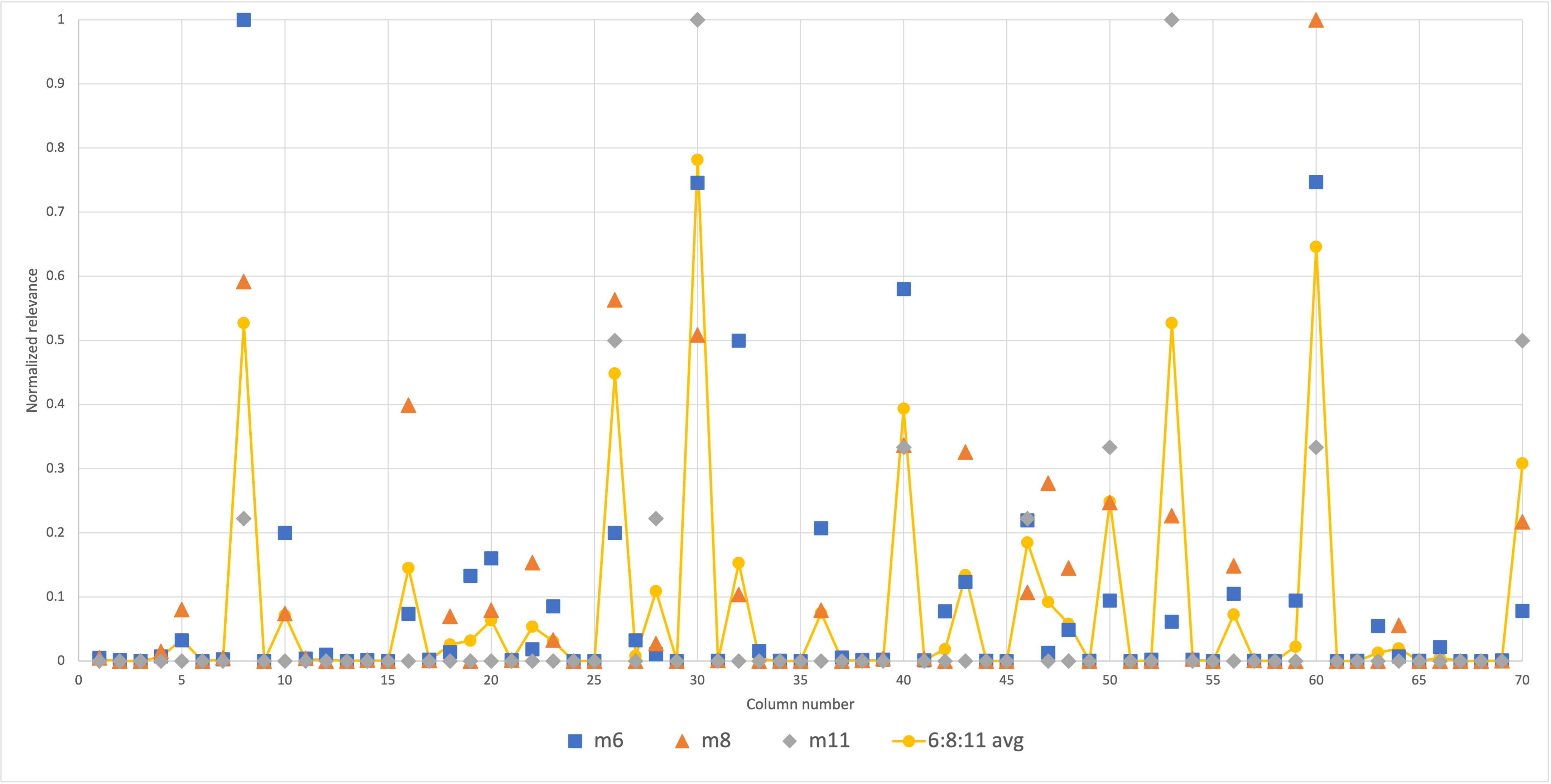}
\caption{0-1 normalized relevance numbers for three minimal support levels}
\label{fig:120ND}
\end{center}
\end{figure}

\begin{figure}[htbp] 
\begin{center}
\includegraphics[scale=0.5]{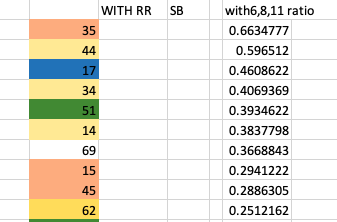}
\caption{Top 10 attributes for a Sea Breeze day}
\label{fig:top10SB}
\end{center}
\end{figure}

As the result, the top 10 attributes with aggregated relevance values, associated with the sea breeze, are given in Figure \ref{fig:top10SB}. Per color code described in section \ref{second stage}, these all have either dark colors or medium colors, and half of attributes are either \emph{high pressure} or \emph{constant dew point} attributes.

\begin{figure}[htbp] 
\begin{center}
\includegraphics[scale=0.5]{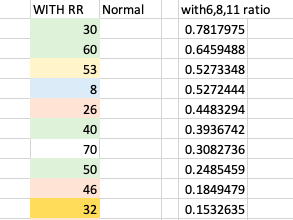}
\caption{Top 10 attributes for a Normal day}
\label{fig:top10ND}
\end{center}
\end{figure}

The top 10 attributes associated with normal day are all non-dark colors, as seen on Figure \ref{fig:top10ND}. \emph{Offshore wind} appears 4 times, for time periods between 11pm and 7am, two others are \emph{low pressure}.

To illustrate the computation, we switch to smaller value of attributes $s=5$,  and show how the formula for $\F(x)$ would be used to predict the sea breeze on June 4, 2017.

\begin{table}[t]
\begin{center}
\begin{tabular}{ |  c| c   | c|} 
  \hline
  $a_i$ & attribute \# & Description \\
  \hline
 $a_1$ & 35 & high pressure 11 pm - 1 am\\
 \hline
$a_2$  & 44 & dew point constant  1-3 am\\
\hline
$a_3$ & 17 & pressure rising 9 pm - 11 pm\\
\hline
$a_4$ & 34 & dew point constant  11 pm - 1 am \\
\hline
$a_5$ & 51 & onshore wind 3-5 am \\
\hline
\end{tabular}
\caption{Sea breeze attributes for an example in $\F$ computation}
\label{Tab6}
\end{center}
\end{table}

The top 5 attributes $a_i$ for the sea breeze and normal day are listed in Tables \ref{Tab6} and \ref{Tab7}, respectively, and the corresponding coefficients $k_i$, are listed in Figures \ref{fig:top10SB} and \ref{fig:top10ND}. 

\begin{table}[t]
\begin{center}
\begin{tabular}{ |  c| c   | c|} 
  \hline
  $a_i$ & attribute \# & Description \\
  \hline
 $a_8$ & 30 & offshore wind  11 pm - 1 am \\
 \hline
$a_9$ & 60 & offshore wind  5-7 am\\
\hline
$a_{10}$ & 53 & dew point falling 3 am - 5 am\\
\hline
$a_{11}$ & 8 & pressure falling 7-9 pm\\
\hline
$a_{12}$ & 26 & low pressure 9 pm- 11 pm \\
\hline
\end{tabular}
\caption{Normal day attributes for an example in $\F$ computation}
\label{Tab7}
\end{center}
\end{table}

Vector $\overline{x}$ corresponding to this entry of the data contains two sub-vectors: $x_{sb}$ for the components corresponding to attributes of the sea breeze 

\[\overline{x_{sb}} = \langle x_{35},x_{44},x_{17},x_{34},x_{51} \rangle = \langle 1, 0, 1,1,0\rangle
\]
and sub-vector $\overline{x_{nor}}$ for the attributes of the normal day:
\[\overline{x_{nor}} = \langle x_{30},x_{60},x_{53},x_{8},x_{26} \rangle = \langle 1,1,0,0,0\rangle\]. 

Thus $\overline{x} = \langle 1, 0, 1,1,0, 1,1,0,0,0 \rangle$. Observations on June 4, 2017 show the presence of three attributes associated with the sea breeze and two associated with normal day predictions, and actual addition and subtraction of corresponding coefficients $k_i$ produce positive number:

\[
\F(\overline{x}) = 0.663478 *1 + 0.596512 *0+ 0.460862 *1 + 0.406937 *1+ 0.393462 *0
\]
\[
\ \ \ \ \ \ \ \ \ 
- 0.781797 *1 - 0.645949*1 - 0.527335 *0- 0.527244 *0 - 0.448329*0  
\]
\[= 1.531277 - 1.427746 = 0.103531 > 0 
\]

Thus, we predict this day to be a sea breeze day. It turns out to be a sea breeze day, indeed.

\vspace{0.3cm}

Changing the number of top attributes used in prediction may result in various success rates.
 When using all the attributes, we correctly predict 78 days out of the 112 non-moist days.

\begin{table}[t]
\begin{center}
\begin{tabular}{ |  c| c |c   | } 
  \hline
  ~ & Predicted SB & Predicted Normal \\ 
  \hline
  Actual SB & 40 & 14 \\ 
  \hline
  Actual Normal & 20 & 38 \\ 
  \hline
\end{tabular}
\caption{Contingency table of sea breeze forecast}
\label{Tab9}
\end{center}
\end{table}

Since 
using all attributes gives the overall best result, we focus more on this choice. The false negatives and false positives are further analyzed through Table \ref{Tab9}.

Other success measurements could also be computed:\\

Sensitivity/Recall = 38/(38+14) = 38/52 = 73.1\% \\

F1 score = $40/(40 + \frac{1}{2}\cdot (14+20)) = 70.2$\% \\

Precision rate = 38/(38+20) = 65.5\% \\

Furthermore, the success rate is further broken down into each year in Table \ref{Tab10}.
\begin{table}[t]
\begin{center}
\begin{tabular}{ |  c| c |  c | c|} 
  \hline
  Year & \# of days predicted & \# of non-moist days & success rate \\ 
  \hline
  2020 & 22 & 30 & 73.3 \% \\ 
  \hline
  2019 &17 & 27 &  63.0\% \\ 
  \hline
  2018  & 21 & 29 & 72.4 \%  \\ 
  \hline
  2017  & 18 & 26 &  69.2\% \\ 
  \hline
\end{tabular}
\caption{Success rate yearly}
\label{Tab10}
\end{center}
\end{table}

Table \ref{Tab8} shows the successful prediction rate on non-moist days for various choice of the number of top attributes.

\begin{table}[t]
\begin{center} 
\begin{tabular}{ |  c| c   | c    |} 
  \hline
  Top \# of attributes utilized & \# of days predicted correctly &  Success rate out of 112 non-moist days  \\
  \hline
 All attributes & 78 & 69.6\% \\
  \hline
 Top 10 & 77 & 68.8\% \\
 \hline
 Top 9 & 75 & 67.0\% \\
 \hline
 Top 8 & 75 & 67.0\% \\
 \hline
Top 7 & 76 & 67.9\% \\
\hline
Top 6 & 75  & 67.0\%  \\ 
\hline
Top 5 & 76 & 67.9\%\\
\hline
Top 4 & 75 & 67.0\%  \\
\hline
Top 3 & 75 & 67.0\%   \\
\hline
Top 2 & 74 & 66.1\% \\
\hline
Top 1 & 73 & 65.2\% \\
\hline
\end{tabular}
\caption{Success rates of top attributes}
\label{Tab8}
\end{center}
\end{table}

\subsection{Computation of Forecasting Values on the random subset of the data}

To explore the prediction method of Forecasting Values demonstrated above, the process was repeated after 30 rows were removed from the data.
The rankings of the Forecasting Values were compared between the two runs to validate the overall consistency and usefulness of the method. The expectation was that the rankings might be slightly different,
but the overall important attributes would be similar. 

The 30 rows were chosen using the \emph{random.sample} function in Python:\\

sorted(random.sample(range(3,123),30)).\\

It selected 30 non-repeating random numbers in the range 3-123, which are numbers for the rows of the $D$-basis entries corresponding to 120 days of observation.

This set of 30 random days included 15 sea breeze days, 2 very moist days and 13 normal days. The distribution resembled the overall distribution of those day categories in the entire data set, which made this randomly selected subset reasonable for random subset testing. 

The entries for those 30 rows were changed to 0, so that they were not taken into consideration in the subsequent $D$-Basis runs. 
This is similar to the procedure of row removal in $D$-Basis built into its functionality.

Since the total number of the entry
dropped from 120 down to 90, the parameter of minimal support in the $D$-Basis was also scaled down in order to preserve the consistency of the testing. The scaling down from minimal support values of 6, 8, 11 for 120 effective rows down to 90 effective rows was as follows:
$$ \frac{6}{120} = \frac{m_1}{90} \rightarrow m_1 = 4.5 \rightarrow  m_1 \approx 4$$
$$ \frac{8}{120} = \frac{m_2}{90} \rightarrow m_2 = 6$$
$$ \frac{11}{120} = \frac{m_3}{90} \rightarrow m_3 = 8.25 \rightarrow m_3 \approx 9$$
For $m_1$, we decided to use the floor function of $4.5$ as its ceiling $5$ is too close to $m_2$. Thus, the values of minimal supports were finalized to 
$4, 6, 9$. 

Since in the testing with full data the
row removal of the very moist days brought to the better results, this removal was done on the subset of 90 days as well.

Then, the same process was repeated to compute the Forecasting Values for that modified input matrix. 
The success rates on non-stormy days when using modified list of top attributes are given in Table \ref{Tab11}.

\begin{table}[t]
\begin{center}
\begin{tabular}{ |  c| c   | c   |} 
  \hline
  Top \# of attributes utilized & \# of days predicted correctly &  Success rate out of 84 non-moist days  \\
  \hline
 All rows & 54 & 64.3\%\\
 \hline
 Top 10 & 54 & 64.3\%\\
 \hline
 Top 9 & 54 & 64.3\%\\
 \hline
 Top 8 & 54 & 64.3\% \\
 \hline
Top 7 & 55 & 65.5\% \\
\hline
Top 6 & 55 & 65.5\% \\
\hline
Top 5 & 54 & 64.3\% \\
\hline
Top 4 & 52 & 61.9\% \\
\hline
Top 3 & 55 & 65.5\% \\
\hline
Top 2 & 52 & 61.9\%\\
\hline
Top 1 & 55 & 65.5\% \\
\hline
\end{tabular}
\caption{Success rate on a random subset of observations}
\label{Tab11}
\end{center}
\end{table}

Similarly, the attributes associated with normal days were still light-colored, see Figure \ref{fig:90ND}, and low pressure became more dominant than offshore wind, compared to full data results on Figure \ref{fig:120ND}.

\begin{figure}[htbp] 
\begin{center}
\includegraphics[scale=0.5]{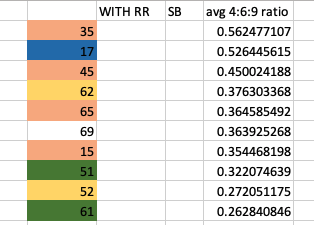}
\caption{Top 10 attributes for a Sea Breeze day on a random subset}
\label{fig:90}
\end{center}
\end{figure}

\begin{figure}[htbp] 
\begin{center}
\includegraphics[scale=0.5]{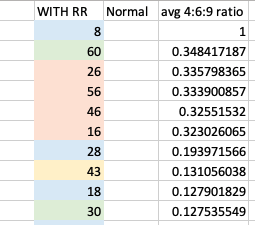}
\caption{Top 10 attributes for a Normal day on a random subset}
\label{fig:90ND}
\end{center}
\end{figure}

It is observed that 'high pressure' 
tends to come up on top of the list for sea breeze as in Figure \ref{fig:90ND}. 4 'high pressure' attributes are now in the top 10 list, including the previous 3 'high pressure' attributes: 35,15,45. Meanwhile, for the normal day list, we still see the 'offshore wind' attributes, and more of the 'low pressure' attributes, including those appearing on the full data: 26 and 46.

\subsection{Comparison with synthetic data}

An earlier study \cite{SCAN18} compared the performance of the $D$-basis algorithm on real data with the random data of similar characteristics. It showed that there was a much lower probability of rules of high support occurring in random data of same size and density, compared to real data. In particular, it showed more uniform distribution of total support figures among all attributes. That result brought the average of relevances across all attributes close to 1, for the wide range of density of random data.

For example, the testing of several thousand tables of size $20\times 32$ in \cite{SCAN18} showed the average relevance values between 1 and 2, i.e., close to insignificant, for the range of densities of entry 1 in the tables between 0.3-0.6. We note that the majority of real data densities does fit into this range, given the rules of conversions between raw data and its binary representation.

To confirm the actual signal from the weather data in our analysis, we ran a similar comparison with random data of similar parameters. The synthetic data was made of 
the unchanged target columns, while the rest of columns were randomized based on original density of entry 1 at 0.402738. The results of three runs of synthetic data on table of size $120\times 72$, targeting column 71, which was not changed compared to real data, are given in Table \ref{Tab12}.

\begin{table}[t]
\begin{center}
\begin{tabular}{ |  c| c |c   | } 
  \hline
  Minimal Support & Synthetic Data (\# of rules) & Real Data (\# of rules) \\ 
  \hline
  6 & 203 & 1035 \\ 
  \hline
  8 & 5 & 266 \\ 
  \hline
  11 & 1 & 33 \\ 
  \hline
\end{tabular}
\caption{Comparison between number or rules in synthetic data and real data}
\label{Tab12}
\end{center}
\end{table}

Since there was only one rule generated for minimal support 11, the relevance values were not computed in random data test.
Thus, only the relevance for minimal supports 6 and 8 were computed and compared with the real data.

\begin{figure}[htbp] 
\begin{center}
\includegraphics[scale=0.4]{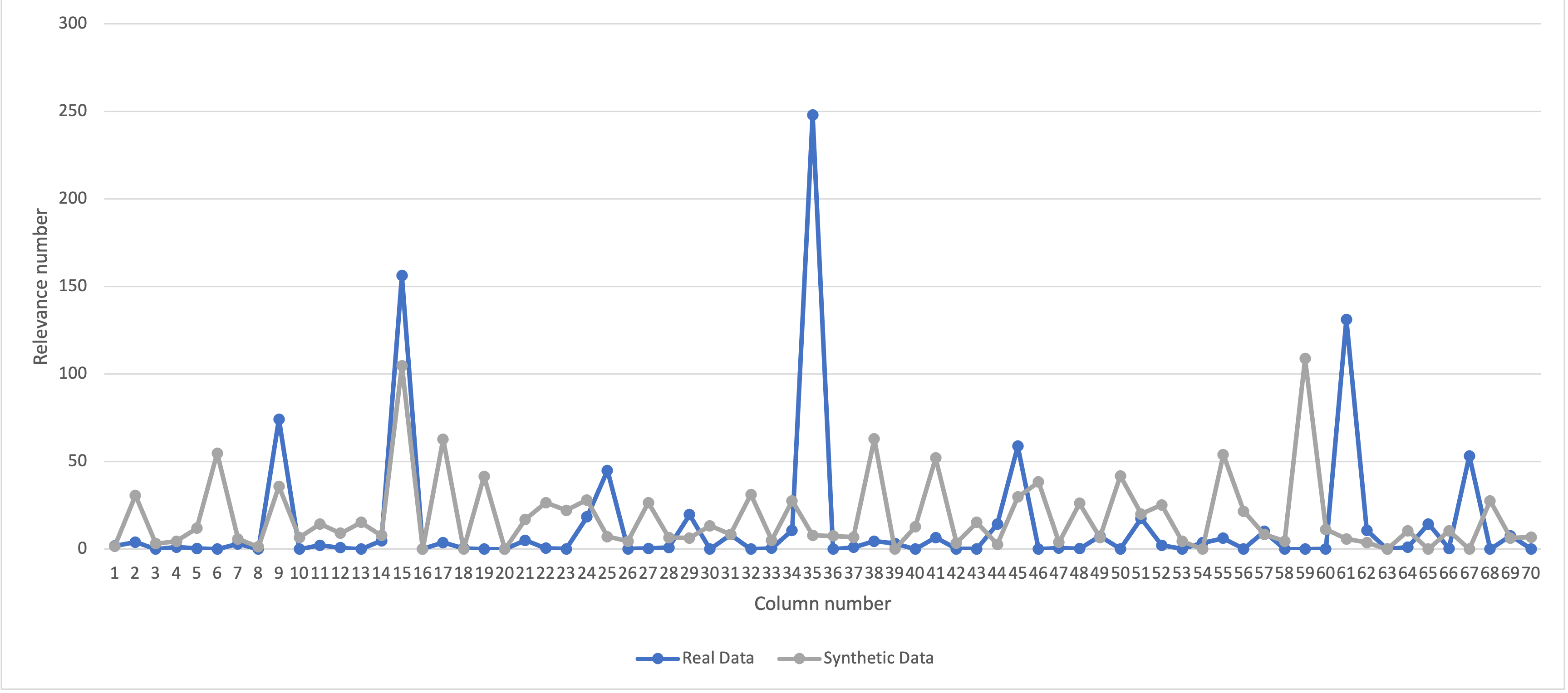}
\caption{Relevance values in real data vs synthetic data in test with minimal support=6}
\label{fig:sup6}
\end{center}
\end{figure}

\begin{figure}[H]
\begin{center}
\includegraphics[scale=0.4]{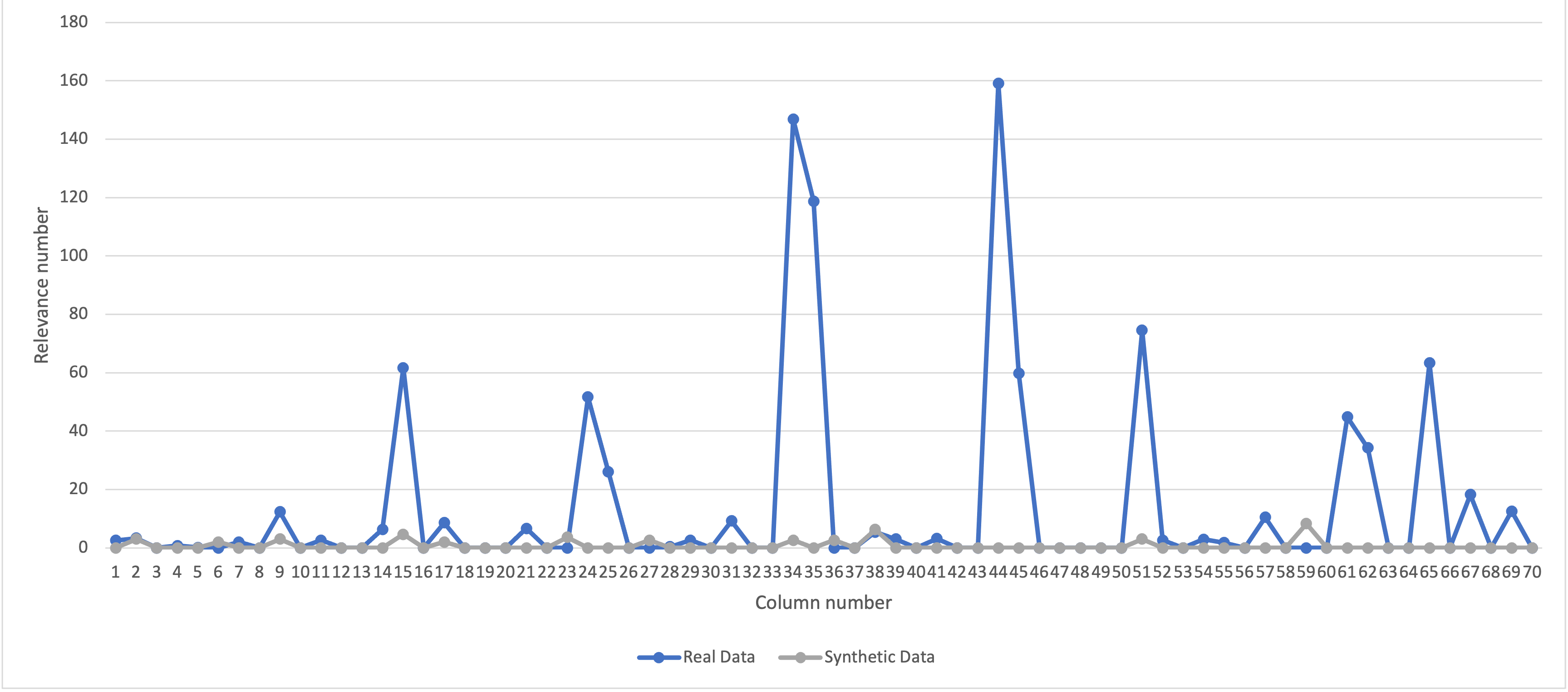}
\caption{Relevance values in real data vs synthetic data in test with minimal support=8}
\label{fig:sup8}
\end{center}
\end{figure}

The histograms on Figures \ref{fig:sup6} and \ref{fig:sup8} show resulting distribution of relevance values $rel_d(a)$ across all attributes $a$ in the data.

The values of $a$ are placed along the $x$-axis, and the component along the $y$-axis is the relevance value computed in the synthetic data for the target column $d=$71. 

For the $D$-basis test with minimal support = 6, the majority of the relevance values were insignificant  and only a few attributes stood out. But as the minimum support increased to 8 and then 11, the number of rules dwindles down considerably and most attributes do not appear in the rules.

In the real data, since each 2-hour time interval has 10 attributes, the peaks in about every 10 attributes are observed; those attributes correspond to the important weather observations occurring in each 2-hour interval. Conversely, the pattern for the synthetic data does not possess any trend as observed by eye and are significantly smaller in magnitude as well.

In the test with the minimal support = 8
significantly less fluctuations in the synthetic data is observed. More importantly, the grey line representing the synthetic data shows considerably smaller total support values, and only 9 attributes have non-zero relevance values, see Figure \ref{fig:sup8}.

In conclusion, from the two figures we observe that the real data demonstrates a considerably strong signal reflected in the high frequency of some attribute appearing in the rules, compared to the flat and uniform frequences of most attributes in the rules, when the ones in the table randomly permuted.

\section{Conclusions and Discussion}

In this study we relied on data collected by weather stations located relatively far inland instead of using the measurement of the sea temperature.
This differs our approach from the standard reliance on temperature gradient between sea and land, when predicting the sea breeze. In the present analysis, we included measurements during the 12 hours overnight prior to the day of prediction, averaging the behavior over two-hour intervals. Thus, our methods could be useful in developing a prediction algorithm for other locations where sea surface temperatures are unavailable. 

To determine the predicted presence or absence of the sea breeze, we used the $D$-basis algorithm. That algorithm retrieves specific implications and association rules between the columns of the binary table, namely, those whose conclusion is a selected column. In our case it was either the column marking sea breeze days occurring during month of June between 2017-2020 on Long Island, New York, or normal days of the same period. The third category of days is very moist, when the synoptic scale features typically control the onshore wind flow, and not the smaller scale sea breeze. The $D$-basis algorithm allows for an analysis in the presence of entries for which the partial failure of implications occurs, thus, very moist days were partially and randomly omitted on multiple runs of the algorithm.

Among the weather observations included in the initial analysis were dynamic parameters: increasing, decreasing, or constant pressure and dew point, as well as the direction of wind in relation to coast line of the south shore of Long Island. Later we added an absolute value of station pressure distinguishing high pressure and low pressure, using a threshold of 29.92'' Hg. Using that definition, high pressure, especially during the first half of the overnight observations between 7pm-1 am, appeared as a prominent predictor of a sea breeze, as well as a constant dew point and early morning (3-5 am) onshore winds. In contrast, a combination of early morning offshore winds, falling or low pressure, and rising dew point appeared as predictors of a normal, non-sea breeze, day. The strongest association was found with observations made between 7 pm and 1 am the previous evening. Those weather variables most closely corresponding with sea breeze formation match the climatological expectations. High pressure typically implies a lack of larger scale synoptic controls on wind direction, such as fronts and cyclones, while a constant dew point also indicates a stable air mass and lack of surface boundaries.

We also determined that by selecting all weather attributes, and converting their frequencies coming from retrieved rules in the $D$-basis algorithm, that we could set a linear combination of binary variables for each column with predetermined coefficients. Each day's observation resulted in values of either 1 or 0 for each binary variable $x_i$, depending on the presence or absence of a particular weather attribute on that day. The total value of this linear combination, when positive, indicates the presence of most sea breeze predictors, and therefore, forecasts a sea breeze day, otherwise, it forecasts a non sea breeze day. Prediction of sea breeze days with this formula was about 70\% accurate: with only 14 wrong predictions over a total 54 sea breeze days. Moreover, the success rate for the year 2020 alone was about 74\%. This success rate is lower than the 85\% rate in \cite{McCabe2023}, which approached the same region and time period. However, that particular study included both 'classic' and 'hybrid' sea breeze days, the latter of which we attempted to eliminate from consideration in the present study, through the synoptic climatology requirement, which means the success rates are not directly comparable. This is due to the fact that the \cite{McCabe2023} study was less restrictive in the sense that days where synoptic influences impacted the sea breeze were counted as sea breeze days and thus an accurate prediction made by their algorithm. These contrasting definitions of a Long Island sea breeze event reflect the notion that there is no universally accepted definition for what constitutes a sea breeze event, and further, that the definition and associated identification algorithm can vary based on its application. For example, when attempting to forecast the effect of enhanced wind speeds from the sea breeze on wind energy potential, the broader definition employed in \cite{McCabe2023} would be more appropriate. Conversely, if attempting to isolate the mesoscale influence of the sea breeze on near-surface weather conditions, a more restrictive definition, such as the one used in the present study, would be necessary.  

It should also be noted that different combinations of predicting attributes are possible too. For example, using only two attributes resulted in a similar success rate in prediction. Moreover, an adjustment of coefficients in the forecasting formula could be undertaken when more observation are collected over time, thus allowing the formula to potentially be more accurate. Last, in our study, we did not distinguish between sea breeze days with respect to the classification of a prior day. Indeed, a simple persistence forecast (i.e., today will be the same as yesterday) can also be somewhat effective when predicting whether a sea breeze will occur.
Nevertheless, the algorithm developed here was still more accurate than a persistence forecast would have been, and future iterations are likely to be even more accurate when a larger sample of data can be ingested, allowing for this tool to have value to weather forecasters in the region.\\


\paragraph{Acknowledgments}
We appreciate the technical support of Dr. Oren Segal in the Hofstra University Department of Computer Science for his involvement in various projects related to the $D$-basis algorithm. All tests and computations were performed using the Virtual Machine at the Hofstra University Fred DeMatteis School of Engineering and Applied Science's Computing Center. We also appreciate the initiative and support of Justin Cabot-Miller, who helped establish this project while an undergraduate research assistant at Hofstra.






\printbibliography 


\begin{appendix}
\section{Appendix: The description of the \emph{D}-basis algorithm}
\subsection{$D$-basis algorithm}
\label{A:Dbas}

In this work we explore an approach in weather forecasting that utilize the discovery of association rules in the binary data. 
Association rules took off as a premiere tool of Data Mining analysis of transaction data since introduction of \emph{Apriori} algorithm in \cite{Agra93}. One of more recent surveys on association rules is in \cite{Balc1o}.

\emph{Apriori} remains a main tool of extraction of association rules in binary data and was included into libraries of R and Microsoft office.

In our work we employ a novel algorithm with the background in Formal Concept Analysis (FCA), see \cite{Wil99}.
It is based on non-trivial mathematical theorem that associates a finite binary table with a uniquely defined \emph{Galois lattice}, see \cite{Birk40,Barb70}. The structure of that mathematical object is fully described by some set of \emph{implications}, defined on the set of columns (or rows) of the given table. Implications can be understood in the framework of Data Mining as association rules of confidence = 1.

The $D$-basis was introduced in \cite{ANR13}, and the theoretical background of the algorithm of the $D$-basis extraction from a binary table is described in \cite{AN17}. It discovers the implications $S \rightarrow d$ in a table with entries $0$ and $1$. Here $S$ is a subset of attributes/columns (weather observation attributes) and $d$ is another column (say, an indicator of sea breeze day). The rows of the table represent the days of the observation. 

The rule $S \rightarrow d$ is found in the table, if in each row (each day of observation in our case) the following is true: if all entries in columns (weather observations) of subset $S$ are $1$, then column $d$ also has entry $1$. 

The \emph{support} of a rule $S \rightarrow d$ is the number of rows (days), where all entries in columns $S\cup d$ are $1$, i.e. where the rules is \emph{validated}.

In practice, the algorithm computes rules that hold in almost all rows of the table, that is, rules such that
the probability of $d=1$, when the entries in $S$ are all $1$, is above a given threshold.
More precisely, parameter \emph{the confidence of rule} $S\rightarrow d$ is computed as follows:
\[
conf (S\rightarrow d) = \frac{\# rows \text{ where all } S \text { and } d \text{ have values 1}}{\# rows \text{ where all } S \text{ have values 1}}
\]
Confidence of rule $S\to d$ is $1$ iff $d=1$ in every row, where all values of $S$ are 1.
In order to retrieve the rules that may fail in some rows, thus, have $conf < 1$, one may remove those rows and retrieve the rules from a sub-table.

This functionality of the application is described in publication \cite{SCAN18}.
\vspace{0.3cm}

{\bf Example 8-1.}
On one of earlier tests when the data of three years was used, thus, there were 90 rows of the table representing 90 days of three months of June between 2017-2019, the $D$-basis was applied with row-removal parameter of the following list of rows:\\

\emph{Row removal 3 of $\{13,14,19,22,33,67,87,90\}$. These are the rows very moist days are used as prediction.}\\

The meaning of such test is that algorithm will run multiple times on sub-tables of the table, when 3 rows of the given list are removed. Since 
very moist days may disturb the weather patterns and may not predict correctly what happens on the following day, 
some of these days are removed from analysis.

In addition to the test above, 
the list of removed rows was also shifted by one:\\

\emph{Row removal 3 of $\{12,13,18,21,32,66,86,89\}$. These are the rows that are predicting very moist days.}\\

This is done because the the days prior to very moist days may also have disturbed weather patterns.\\

\subsection{Ranking attributes relevant to  a target attribute}\label{A:rank}

Unlike the most approaches in mining the association rules, when some rules are selected based on some techniques measuring the rules themselves, we are looking at the measurement of all attributes as being more or less \emph{relevant} to some chosen \emph{target} attribute.

To this aim, the $D$-basis algorithm is most suitable, since it may retrieve only a sector of the total basis of implications describing the table, namely, the rules $S\to d$, when $d$ is some fixed \emph{target} attribute.

Instead of looking into the rules themselves, we measure the frequency of any other attribute $a$ as it appears in the antecedents of the rules $S\to d$, together with other attributes. Note that the same idea is highlighted in \cite{Coul21}, in relation to 18 rules considered there, which  connect several weather attributes with the same target='heavy rain'.

Our parameter of \emph{the relevance}, which appeared first in \cite{ANO15}, when applying the $D$-basis to the medical data, requires the computation of the rules not only on target $d$ but also on its complement $\neg d$, which may or may not appear in original data.

Another advantage of our approach is that we do not shy away from having a big amount of retrieved rules, because they provide better representation of all attributes and allow better comparison of attributes related to a given target. 

The top attributes were identified through testing with variation of the \emph{minimal support}, which refers to percentage of observations validating the rules connecting attributes and sea breeze/absence of sea breeze. In our testing, we varied minimal support between 5\% and 15\% of observations in the testing set.

\subsection{Formula for the parameter of relevance}\label{A:relevance}

For a fixed column $d$ and any other column $a$, one can compute the total support of all rules $S \rightarrow d$ such that $a$ is in $S$. This parameter shows the frequency that $a$ appears in implications targeting $d$. The algorithm can also compute a similar frequency of $a$, when targeting $\neg d$, i.e., an additional column where all entries in $d$ are switched. The ratio of the two frequencies gives the \emph{relevance} of attribute $a$ to $d$. Thus, all attributes of the table can be ranked in their relevance to a fixed attribute $d$.

The $D$-basis can be applied to the entry table formed by attributes available in objects/instances of observation, and choose $d$ as a marker for a particular property/attribute (in our case - for the sea breeze, or normal day). It was used in ovarian cancer analysis in \cite{ANO15}, and in stomach cancer analysis in \cite{NCSLA21}, where observations were represented by patients in the study, and target attribute was an indicator of long survival after the treatment.

Let us give more precise definition how the relevance of attribute $a$ with respect to target attribute $d$ is computed. 
For each attribute $a \in S\setminus{d}$, the important parameter of relevance of this attribute to $d \in X$ is a parameter of \emph{total support}, computed with respect to any set of rules/basis $\beta$ of association rules describing the table (in our case, it is portion of the $D$-basis which only includes rules of requested minimum-support at least $\delta$):
\[
\tsup_d(a)=\Sigma \{\frac{|sup(X)|}{|X|}*conf(X\to d):  a\in X, (X\rightarrow d)\in \beta \text{ and } |sup(X)|\geq\delta\} .
\]

Thus $\tsup_d(a)$ shows the frequency of parameter $a$ appearing together with some other attributes in implications $S\rightarrow d$ of the basis $\beta$. The contribution of each implication $S\rightarrow d$, where $a \in S$, into the computation of total support of $a$ is higher when the support of $X$ is higher, i.e., column $a$ is marked by $1$ in more rows of the table, together with other attributes from $X$, but also when $X$ has fewer other attributes besides $a$.\\

{\bf Example 8-2.} In the earlier stage of testing, when we used 58 attributes (columns) on 120 days representing rows, we targeted column $d=55$: \\

55: Next Day Sea Breeze (1=SB, 0=Normal or very moist)\\

and one of about 500 implications with minimum support $\geq 6$ that appeared in the output may look as follows:\\

5 16 30 52 $\to$ 55 ; RealSupport = 7; rows = 23, 62, 78, 80, 109, 110, 118, \\
		conf=0.78, fail rows = 18, 66\\

Here the attributes on the left side of implication mean:\\
5: SSC Moist Day (1=yes, 0=no)\\
16: Pressure Falling between 9-11 pm (1=yes, 0=no)\\
30: Dew Point Consistent (1=yes, 0=no) within (-1,1) between 11 pm -1 am\\
52: Dew Point Rising (1=yes, 0=no) between 5-7 am\\

List of rows for the RealSupport means that on each of days 23, 62, 78, 80, 109, 110 and 118 in our data of 120 days it was a sea breeze day (1 in column 55) and all four attributes appeared. The confidence is $0.78 = 7/9$, because all four attributes appeared on 9 days, but on days 18 and 66 attribute 55 did not appear, since it was not a sea breeze.  

For each of attributes $a=$ 5, 16, 30, 52 this implication produces the following contribution into $tsup_{55}(a)$ : $\frac{7}{4}*0.78=1.36$. The totals of $tsup_{55}(a)$ are then combined across all implications, where $a$ appears. For example, $tsup_{55}(5)=34.27$ and $tsup_{55}(30)=136.36$.\\
\\

While the frequent appearance of a particular attribute $a$ in implications $S\rightarrow d$ might indicate the relevance of $a$ to $d$, the same attribute may appear in implications $U\rightarrow \neg d$. The attribute $\neg d$ may not be present in the table and can be obtained by converting the column of attribute $d$ into its complement. \\

Let $\beta (\neg d)$ be the basis of closure system obtained after replacing the original column of attribute $d$ by its complement column $\neg d$. Then the \emph{total support} of $\neg d$ can be computed, for each $a \in X\setminus d$, as before:
\[
\tsup_{\neg d}(a)=\Sigma \{\frac{|sup(X)|}{|X|}* conf(X\to \neg d):  a\in X, (X\rightarrow \neg d)\in \beta(\neg d)\} .
\]

{\bf Example 8-2.} (continued)

For the test described in Example above, we also run $D$-basis with the target column 56, which is marked by 1 for every \emph{non-sea breeze} day\\

56: Complement of 55 (1=Normal or very moist, 0=SB)\\

For example, the same attribute 5 appeared in one of implications in the output:\\

5 18 26 54 $\to$ 56 ; Real Support = 6; rows = 49, 55, 59, 60, 103, 120, conf=1.00\\

As the result, $tsup_{56}(5)$ will have an addend for this implication: $\frac{6}{4}*1=1.5$. Computing the sum across all implications that have attribute 5, we will get $tsup_{56}(5)=114.88$. According to this number, attribute 5 appears more frequently in all association rules of minimal support $\geq 6$ on \emph{non-sea breeze} days.\\

Define now the parameter of relevance of parameter $a \in X\setminus d$ to parameter $d$, with respect to basis $\beta$: 
\[
\rel_d(a) = \frac{\tsup_d (a)}{\tsup_{\neg d} (a) +1}.
\]

The highest relevance of $a$ is achieved by a combination of high total support of $a$ in implications $S\rightarrow d$ and low total support in implications $U\rightarrow \neg d$.
This parameter provides the ranking of all parameters $a \in X\setminus d$.\\

{\bf Example 8-2.} (continued) Using two numbers $tsup$ computed for attribute $a=5$, when targeting $d=55$, we come up with relevance of $5$ for the sea-breeze:
\[
\rel_{55}(5) = \frac{\tsup_{55} (5)}{\tsup_{56} (5) +1}=\frac{34.27}{114.88+1}\approx 0.29
\]

The relevance number smaller than 1 indicates that attribute appears more frequently on the days that are \emph{not sea-breeze} than on sea breeze days.\\

More precisely, the relevance is compared with number $\alpha=\frac{sup(55)}{sup(56)}$, which is the ratio of ones to zeroes in the target column. In our case, it is approximately 1.

\end{appendix}


\end{document}